\documentclass[preprint,aps,floatfix,onecolumn]{revtex4}
\usepackage{epsfig,amsmath,bbm,graphicx,color}

\begin{document}

\title{New force replica exchange method and
protein folding pathways probed by force-clamp technique}

\author{ Maksim Kouza$^1$, Chin-Kun Hu$^{2,3}$ and Mai Suan Li$^{1,\star}$}

\address{$^1$Institute of Physics, Polish Academy of Sciences,
Al. Lotnikow 32/46, 02-668 Warsaw, Poland\\
$^2$Institute of Physics, Academia Sinica, Nankang,
Taipei 11529, Taiwan\\
$^3$Center for Nonlinear and Complex Systems and Department of
Physics, Chung Yuan Christian University, Chungli 32023, Taiwan\\
$^{\star}$ Corresponding author: masli@ifpan.edu.pl}

\begin{abstract}
We have developed a new extended replica exchange method to study
thermodynamics of a system in the presence of external force.
Our idea is based on the exchange between different force replicas
to accelerate the equilibrium process. This new approach was applied to
obtain the force-temperature phase diagram and other thermodynamical
quantities of the three-domain ubiquitin.
Using the C$_{\alpha}$-Go model and the Langevin dynamics, we have shown that
the refolding pathways of single ubiquitin depend on which terminus is fixed.
If the N-end is fixed then the folding pathways are different compared to the
case when both termini are free, but fixing the C-terminal does not change
them. Surprisingly, we have found that the anchoring terminal
does not affect the pathways of individual secondary structures of
three-domain ubiquitin, indicating the important role
of the multi-domain construction. Therefore,
force-clamp experiments, in which one end of a protein is kept fixed,
can probe the refolding pathways of a single free-end ubiquitin if one uses
either the poly-ubiquitin or a single domain with the C-terminus anchored.
However, it is shown that anchoring one end does not affect refolding
pathways of the titin domain I27, and the force-clamp spectroscopy is
always capable
to predict folding sequencing of this protein.
We have obtained the reasonable estimate for unfolding barrier of
ubiqutin,using the microscopic theory for the dependence of unfolding time on
the external force.
The linkage between residue Lys48 and the C-terminal of ubiquitin
is found to have
the dramatic effect on the location of the transition state
along the end-to-end distance reaction coordinate, but the multi-domain
construction leaves the transition state almost unchanged.
We have found that the maximum force in the
force-extension profile from constant velocity force pulling simulations
depends on temperature nonlinearly. However, for some narrow
temperature interval this dependence becomes linear, as
have been observed in recent experiments.
\end{abstract}

\maketitle
\section{Introduction}

Protein ubiquitin (Ub) continues to attract the attention of researchers because
there exist many processes in living systems where it plays
the vital role. Usually, Ub presents  in the form of a polyubiquitin chain
 that is conjugated to other proteins.
Different Ub linkages lead to  different biological functions.
In  case of Lys48-C and N-C linkages  polyubiquitin chain serves as a signal for
degradation  proteins \cite{Thrower_EMBO2000, Kirisako_EMBO2006},  whereas
in the Lys63-C case it plays completely different functions, including
DNA repair,
polysome stability and endocytosis
 \cite{Hofmann_Cell1999, Spence_Cell2000, Galan_EMBO1997}.

The folding properties of Ub have been studied in detail by experiments
\cite{Went_PEDS05,Sosnick_ChemRev06} as well
as by simulations \cite{Fernandez_JCP01,Sorenson_Proteins02}.
The unfolding of Ub under thermal fluctuations was investigated experimentally
by Cordier and Grzesiek \cite{Cordier_JMB02} and by Chung {\em et al.}
\cite{Chung_PNAS05}, and studied
theoretically \cite{Alonso_ProSci98,Irback_Proteins06,MSLi_BJ07}.
There is  strong evidence that thermal unfolding pathways are reverse
of the folding ones.
The mechanical unfolding of Ub has been studied by several groups
\cite{Carrion-Vazquez_NSB03,Schlierf_PNAS04,West_BJ06,Irback_PNAS05}.
It was found that the unfolding may proceed through rare intermediates,
but the overall behavior remains two-state, although a three-state
scenario was also reported \cite{Fernandez_NaturePhys06}. The mechanical
unfolding pathways
are very different from the thermal pathways, and depend on what
terminal is kept fixed \cite{MSLi_BJ07}.

Recently Fernandez and coworkers \cite{Fernandez_Sci04} have applied
the force-clamp technique to probe refolding of Ub under quench
force, $f_q$, which is smaller than the equilibrium critical force
separating
the folded and unfolded states. This kind of experiment
has a number of advantages
over the standard ones. In the force-clamp technique, one can
control starting conformations  which are well prepared by applying
the large initial
force of several hundreds of pN. Moreover, since the quench force slows down the
folding process, it is easier to monitor refolding pathways.
However, this begs the important question as to whether the experiments with
one end of the protein anchored probes the same folding pathways as a free-end
protein.
Recently, using a simple Go-like model, it has been shown that
fixing the N-terminal of Ub changes its
folding pathways
\cite{Szymczak_JCP06}. If it is so, the force-clamp technique
in which the N-terminal is anchored is not useful
for prediction of folding pathways of the free-end Ub.

When one studies thermodynamics of a large system like multi-domain
ubiquitin the problem of slow dynamics occurs,
due to the rough free energy landscape.
This
problem might be remedied using
the standard replica exchange method in
the temperature space in the absence of external force
 \cite{Hukushima_JPSC96,Sugita_ChemPhysLett99,Phuong_Proteins05}
as well as in the presence of it \cite{Li_JPCB06}.
% However,
% a similar approach  is not available for systems subject to external force.
However,
if one wants to construct the force-temperature phase diagram,
then this approach becomes inconvenient because one has to collect
data at different values of forces.
Moreover, the external force increases unfolding barriers and a system may
get trapped in some local minima. In order to have better sampling for a system
subject to external force we propose a new replica exchange method in which
the exchange is carried not in the temperature space but in the force space,
 i.e.
the exchange between different force values. This procedure would help
the system to escape from local minima efficiently.

In this paper we address three topics. First, we develop
a new version of the replica exchange method to study thermodynamics of a
large system under the force. The basic idea is that for
a given temperature we perform simulation at
different values of force and the exchange between them is carried out according
to the Metropolis rule.
This new approach has been employed to obtain the force-temperature phase
diagram of the three-domain ubiquitin. Within our choice
of force replicas it speeds up computation about four times
compared to the conventional simulation.
Second, question we ask
is
under what conditions the force-clamp technique still gives
the same folding pathways as for the free-end Ub.
%Third, is the force-clamp spectroscopy capable to predict the refolding
%sequence for other proteins like the well-studied titin domain I27?
Third, we determine the temperature dependence of the unfolding
force, the unfolding barriers and
the location of the transition state
of the single Ub  and of the three-domain Ub
(the three-domain Ub will be also referred to as trimer).

Because the study of
refolding of 76-residue Ub
by all-atom simulations is beyond
present computational facilities, the Go modeling is an
appropriate choice.
The Go-like model \cite{Clementi_JMB00} was proved
\cite{MSLi_PNAS06,MSLi_BJ07} to give not only
qualitative but also quasi-quantitative agreement with existing
refolding and unfolding experiments. Therefore we will use it to study
the folding and unfolding of a single and multi-domain Ub.
In agreement with an earlier study \cite{Szymczak_JCP06},
we show that fixing
N-terminal of the single Ub changes its folding pathways. Our new finding
is that
anchoring
C-terminal leaves them unchanged. More importantly,
we have found that for the three-domain Ub
with either end fixed,
each domain follows the same folding pathways as for the free-end single
domain. Therefore,
to probe the folding pathways of Ub by the
force-clamp technique one can either use the single domain with C-terminal
fixed, or several domains with either end fixed.
In order to check if the effect of fixing one terminus is valid for other
proteins, we have studied  the titin domain I27.
It turns out
that the fixation of one end of a polypeptide chain
does not change the refolding pathways of I27.
 Therefore the
force-clamp can always predict the refolding pathways of the single as well as
multi-domain I27. Our study suggests that the effect of the end fixation
is not universal for all proteins, and the force-clamp
spectroscopy should be applied with caution.

The third part of this paper was inspired by the recent
pulling experiments of Yang {\em et al.} \cite{Yang_RSI06}.
Performing the
experiments in the temperature interval between 278 and
318 K, they found that the unfolding
force (maximum force in the force-extension profile), $f_u$,
of Ub  depends on temperature
linearly. In addition, the corresponding slopes of the linear behavior
have been found to be
independent of pulling velocities.
An interesting question that arises is if the linear dependence
of $f_u$ on $T$ is valid for this
particular region, or it holds for the whole temperature interval.
Using the same Go model \cite{Clementi_JMB00}, we can reproduce the
experimental results of Yang {\em et al.} \cite{Yang_RSI06}
 on the quasi-quantitative level.
More importantly, we have shown that for the entire temperature
interval the dependence is not linear, because a protein is not an
entropic spring in the temperature regime studied.

As a part of the third topic, we have
 studied the effect of multi-domain construction
 and linkage
on the location
of the transition state along the end-to-end distance reaction
coordinate, $x_u$.
It is found that the multi-domain construction has a minor effect on $x_u$
but,
in agreement with the experiments
\cite{Carrion-Vazquez_NSB03}, the Lys48-C linkage has
the strong effect on it.
Using the microscopic theory for unfolding dynamics \cite{Dudko_PRL06},
we have determined the unfolding barrier for Ub.

To summarize, in this article we have obtained the following novel results.
We have proposed a new replica exchange method, in which the exchange
is carried out between different external forces at a given temperature.
It is shown that
the force-clamp technique probes the same folding pathways as the
free-end Ub, if one keeps the C-terminal fixed or uses
the multi-domain Ub with what ever terminal anchored.
Contrary to the $\alpha/\beta$-protein ubiquitin, anchoring
one end has a minor effect on refolding pathways of the titin domain
I27. We attribute this to its $\beta$-sandwich structure, which is
mechanically more stable.
If Ub is pulled at Lys48 and C then, in the Bell approximation,
we  obtained the
parameter $x_u \approx$ 0.61 nm.
This estimate agrees well with the experimental result
of Fernandez group \cite{Carrion-Vazquez_NSB03}.
We propose to fit simulation data with different theoretical schemes
to determine the distance between the transition state and the native state,
$x_u$, for biomolecules.
Our estimate for unfolding barrier of Ub is in reasonable agreement
with the experiments.
It has been demonstrated that the temperature dependence of the unfolding force
is linear for
some narrow interval, but that
this dependence should be nonlinear in general.

\section{Model}

\begin{figure}
\epsfxsize=6.3in
\vspace{0.2in}
\centerline{\epsffile{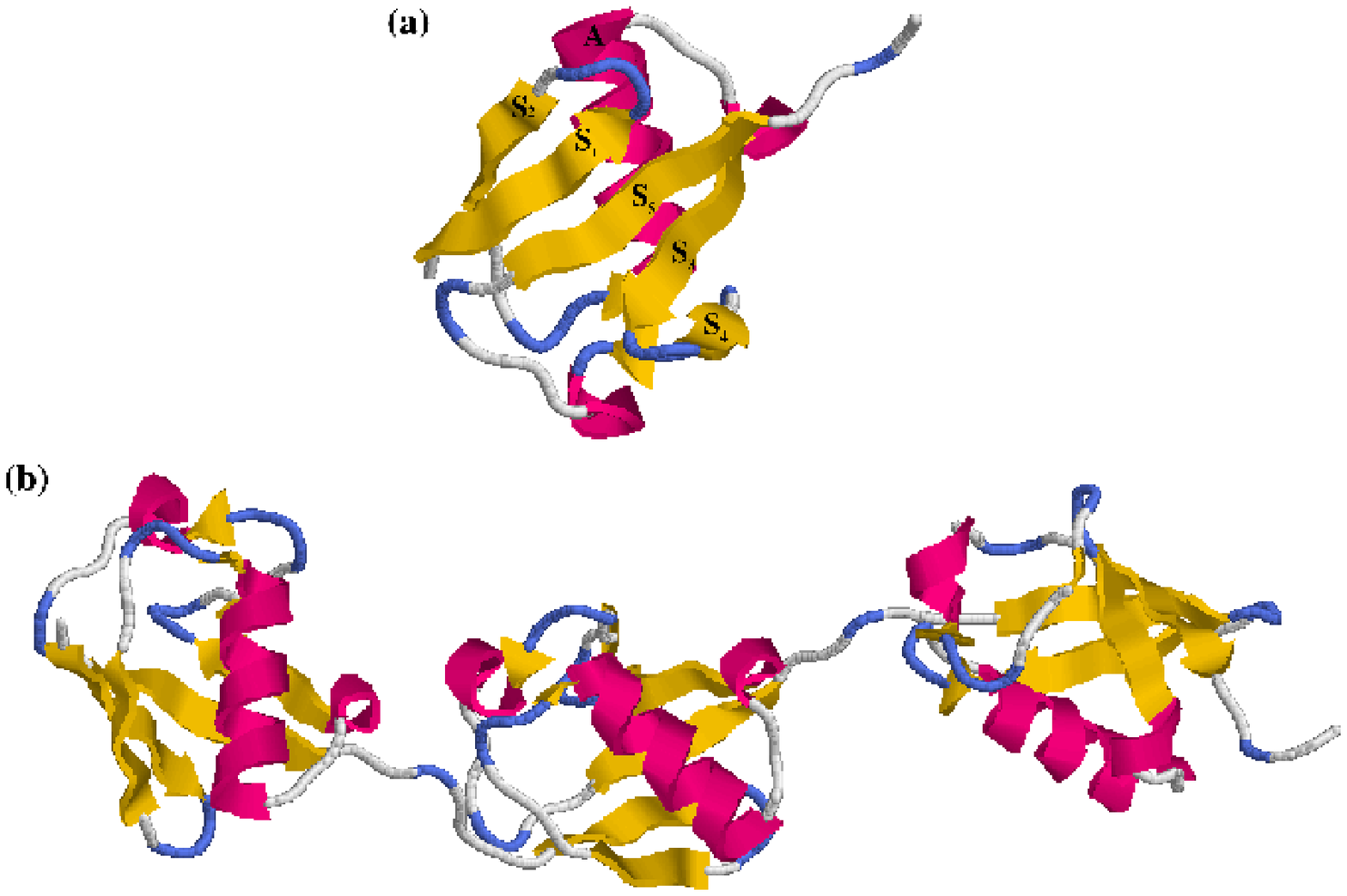}}
\caption{}
\label{native_structure}
\end{figure}

{\em Three-domain model of ubiquitin.} Since the native
conformation
of poly-ubiquitin is not available yet, we have
to construct it for Go modeling. The native conformation
of single Ub, which has five $\beta$-strands (S1 - S5) and
one helix (A), was taken from the PDB (PI: 1ubq,
Fig. \ref{native_structure}a).
We assume that
residues $i$ and $j$ are in native contact if
$r_{0ij}$ is less than a cutoff distance $d_c$ taken to be $d_c =
6.5$ \AA, where $r_{0ij}$ is the distance between the residues in
the native conformation. Then the single Ub has 99 native contacts.
We constructed the three-domain Ub
(Fig. \ref{native_structure}b) by translating one unit by the distance
$a =3.82 \AA \;$ along the end-to-end vector,
and slightly rotating it to have 9
inter-domain contacts (about 10\% of the intra-domain contacts).

{\em The Go-like modeling}. The energy of the Go-like model
for the single as well as multi-domain Ub is as follows \cite{Clementi_JMB00}
\begin{eqnarray}
E \; &=& \; \sum_{bonds} K_r (r_i - r_{0i})^2 + \sum_{angles}
K_{\theta} (\theta_i - \theta_{0i})^2 \nonumber \\
&+& \sum_{dihedral} \{ K_{\phi}^{(1)} [1 - \cos (\phi_i -
\phi_{0i})] +  K_{\phi}^{(3)} [1 - \cos 3(\phi_i - \phi_{0i})] \}
\nonumber \\
& + &\sum_{i>j-3}^{NC}  \epsilon_H \left[ 5\left(
\frac{r_{0ij}}{r_{ij}} \right)^{12} - 6 \left(
\frac{r_{0ij}}{r_{ij}}\right)^{10}\right] + \sum_{i>j-3}^{NNC}
\epsilon_H \left(\frac{C}{r_{ij}}\right)^{12}
. \label{Hamiltonian}
\end{eqnarray}
Here $\Delta \phi_i=\phi_i - \phi_{0i}$,
$r_{i,i+1}$ is the distance between beads $i$ and $i+1$, $\theta_i$
is the bond angle
 between bonds $(i-1)$ and $i$,
$\phi_i$ is the dihedral angle around the $i$th bond and
$r_{ij}$ is the distance between the $i$th and $j$th residues.
Subscripts ``0'', ``NC'' and ``NNC'' refer to the native
conformation, native contacts and non-native contacts,
respectively.

The first harmonic term in Eq. (\ref{Hamiltonian})
accounts for chain
connectivity and the second term represents the bond angle potential.
The potential for the
dihedral angle degrees of freedom is given by the third term in
Eq. (\ref{Hamiltonian}). The interaction energy between residues that are
separated by at least 3 beads is given by 10-12 Lennard-Jones potential.
A soft sphere repulsive potential
(the fourth term in Eq. \ref{Hamiltonian})
disfavors the formation of non-native contacts.
We choose $K_r =
100 \epsilon _H/\AA^2$, $K_{\theta} = 20 \epsilon _H/rad^2,
 K_{\phi}^{(1)} = \epsilon _H$, and
$K_{\phi}^{(3)} = 0.5 \epsilon _H$, where $\epsilon_H$ is the
characteristic hydrogen bond energy and $C = 4$ \AA.

Since for the trimer
$T_F=0.64 \epsilon_H/k_B$ (see below), which is very close
to $T_F=0.67 \epsilon_H/k_B$ obtained for the single
Ub by the same model \cite{MSLi_BJ07}, we will use the same
energy unit $\epsilon_H= 4.1~{\rm kJ/mol}$ as in our previous work.
This unit was obtained by equating the simulated value
of $T_F$ to the experimental
$T_F = 332.5 K$
\cite{Thomas_PNAS01}.
The force
unit is then $[f] = \epsilon_H/\AA \, = 68.0$ pN.
Most of our simulations were performed at temperature
$T=285$ K = $0.53 \epsilon _H/k_B$.

The dynamics of the system is obtained by integrating the following Langevin
equation \cite{Allen_book,Kouza_BJ05}
\begin{equation}
m\frac{d^2\vec{r}}{dt^2} \; \; = \; \;
- \zeta \frac{d\vec{r}}{dt} + \vec{F}_c + \vec{\Gamma},
\label{DynaEq_eq}
\end{equation}
where $m$ is the mass of a bead, $\zeta$ is the friction coefficient,
$\vec{F}_c = -dE/d\vec{r}$. The random force $\vec{\Gamma}$ is a white noise,
i.e. $<\Gamma (t) \Gamma (t')> = 2\zeta k_BT\delta(t-t')$.
It should be noted that the folding thermodynamics
does not depend on the enviroment viscosity (or on $\zeta$)
but the folding kinetics depends
on it. Most of our simulations (if not stated otherwise)
were performed at the friction
$\zeta = 2\frac{m}{\tau_L}$, where the folding is fast.
Here $\tau_L = (ma^2/\epsilon_H)^{1/2} \approx 3$ ps. The
 equations of motion
were integrated using the velocity form
of the Verlet algorithm \cite{Swope_JCP82}
with the time step $\Delta t = 0.005 \tau_L$. In order to check
the robustness
of our predictions for refolding pathways, limited computations
were carried out
for the friction $\zeta = 50\frac{m}{\tau_L}$ which is believed
to correspond to the viscosity of water \cite{Veitshans_FD97}).
In this overdamped limit we use the Euler method for integration
and the time step $\Delta t = 0.1 \tau_L$.

In the constant force simulations, we add an energy
$-\vec{f}.\vec{R}$ to the total energy of the system (Eq. \ref{Hamiltonian}),
where $R$ is the end-to-end distance and $f$ is the force applied
to the both termini or to one of them.
We define the unfolding time, $\tau_U$,
as the average of first passage times to reach a rod conformation.
Different trajectories start from the same native
conformation but, with different random number seeds.
In order
to get the reasonable estimate for $\tau_U$,
for each value of $f$ we have generated 30 - 50 trajectories.

In order to probe folding pathways, for $i$-th trajectory
we introduce the progressive variable $\delta _i =
t/\tau^i_{U}$, where $\tau^i_{U}$ is the unfolding time \cite{MSLi_BJ07}.
Then one
can average the fraction of native contacts over many trajectories
in a unique time window
$0 \le \delta _i \le 1$ and monitor the folding sequencing with
the  help of the progressive variable $\delta$.

In the constant velocity force simulation, we fix the N-terminal and pull the
C-terminal by force $f = K_r(\nu t -r)$,
 where $r$ is the displacement of the pulled atom from its original position
\cite{Lu_BJ98} and the spring constant  $K_r$ is set to be the same as in
Eq. (\ref{Hamiltonian}).
The pulling direction was chosen along the vector from fixed atom to
pulled atom. The pulling speeds are set equal
$\nu = 3.6\times 10^7$ nm/s and 4.55 $\times 10^8$ nm/s which are
about 5 - 6 orders of magnitude faster than those used in experiments
\cite{Yang_RSI06}.

\section{New force replica exchange method and its application}

\subsection{Force replica exchange method}

The equilibration of long peptides at low temperatures is a computationally
expensive job. In order to speed up computation of thermodynamic quantities
we extend the standard replica exchange
method (with replicas at different temperatues)
developed for spin \cite{Hukushima_JPSC96} and peptide systems
\cite{Sugita_ChemPhysLett99} to the case when the replica exchange is
performed between states with different values of
 the external force $\lbrace f_i \rbrace$.
Suppose for a given temperature
we have $M$ replicas $\lbrace x_i, f_i\rbrace$, where
$\lbrace x_i \rbrace$ denotes coordinates and velocities of
residues. Then the statistical
sum of the extended ensemble is
\begin{eqnarray}
Z \; = \; \int \ldots \int dx_1 \ldots dx_M \exp(- \sum_{i=1}^M\beta H(x_i)) = \prod_{i=1}^MZ(f_i).
\label{Z_total_eq}
\end{eqnarray}
The total distribution function has the following form
\begin{eqnarray}
P(\lbrace x,f\rbrace) &=& \prod_{i=1}^M P_{eq}(x_i,f_i), \nonumber\\
P_{eq}(x,f) &=& Z^{-1}(f)\exp(-\beta H(x,f)).
\label{P_total_eq}
\end{eqnarray}
For a Markov process the detailed balance condition reads as:
\begin{eqnarray}
P(\ldots, x_m f_m, \ldots, x_n f_n, \ldots) W(x_m f_m \vert x_n f_n)
 = P(\ldots, x_n f_m, \ldots, x_m f_n, \ldots) W(x_n f_m \vert x_m f_n),
\label{Markov_eq}
\end{eqnarray}
where $W(x_m f_m \vert x_n f_n)$ is the rate of transition
$\lbrace x_m, f_m \rbrace \rightarrow \lbrace x_n, f_n \rbrace$.
Using
\begin{eqnarray}
H(x,f) = H_0(x) - \vec{f}\vec{R} ,
\end{eqnarray}
and Eq. \ref{Markov_eq}
we obtain
\begin{eqnarray}
\frac{W(x_m f_m \vert x_n f_n)}{W(x_n f_m \vert x_m f_n)} \; = \;
\frac{P(\ldots, x_m f_m, \ldots, x_n f_n, \ldots)}{P(\ldots, x_n f_m,
\ldots, x_m f_n, \ldots)} \; = \\ \nonumber \; \frac{\exp[-\beta(H_0(x_n) -
\vec{f}_m\vec{R}_n)
- \beta(H_0(x_m) - \vec{f}_n\vec{R}_m)]}{\exp[-\beta(H_0(x_m) -
\vec{f}_m\vec{R}_m)
- \beta(H_0(x_n) - \vec{f}_n\vec{R}_n)]} \; = \; \exp(-\Delta),
\end{eqnarray}
with
\begin{eqnarray}
\Delta &=& \beta (\vec{f}_m - \vec{f}_n) (\vec{R}_m - \vec{R}_n).
\label{Delta_eq}
\end{eqnarray}
This gives us the following Metropolis rule for accepting or rejecting
the exchange between replicas $f_n$ and $f_m$:
\begin{eqnarray}
W(x f_m | x' f_n) = \left\{ \begin{array}{ll}
         1&, \qquad \mbox{$\Delta < 0$}\\
        \exp(-\Delta)&, \qquad \mbox{$\Delta > 0$}\end{array} \right.
\label{Metropolis_eq}
\end{eqnarray}

\subsection{Application of the force replica exchange method to
construction of the temperature-force phase diagram of three-domain ubiquitin}

Since the three-domain Ub is rather
long peptide (228 residues), we apply the replica exchange method to
obtain its temperature-force phase diagram.
We have performed two sets of the RE simulations. In the first set we fixed
$f=0$ and the RE is carried out in the standard temperature replica space
\cite{Sugita_ChemPhysLett99}, where
12 values of $T$ were chosen in the interval $\left[0.46, 0.82\right]$
in such a way that the RE acceptance ratio was 15-33\%.
This procedure speeds up the equilibration of our system nearly
ten-fold compared to the standard computation without the use of RE.

In the second set, the RE simulation was performed in the force replica
space at $T=0.53$ using the Metropolis rule given by Eq. \ref{Metropolis_eq}.
We have also used 12 replicas with different values of $f$
in the interval $0 \leq f \leq 0.6$ to get
the acceptance ratio about 12\%.
Even for this modest acceptance rate our new RE scheme accelerates
the equilibration
of the three-domain ubiquitin about four-fold. One can expect better
performance by increasing the number of replicas.
However, within our computational facilities we were restricted to
parallel runs on 12 processors for 12 replicas.
The system was equilibrated
during first 10$^5 \tau_L$, after which histograms for the energy, the native
contacts and end-to-end distances were collected
for $4\times 10^5 \tau_L$ .
For each replica, we have generated 25 independent trajectories for
thermal averaging.
Using the data from two sets of the RE simulations and the
extended reweighting technique \cite{Ferrenberg_PRL89} in the
temperature and force space
\cite{Klimov_JPCB01} we obtained the $T-f$ phase diagram and the
thermodynamic quantities of the trimer.

%\section{Results}

%{\bf Temperature-force phase diagram of three-domain ubiquitin}

\begin{figure}
\epsfxsize=6.3in
\vspace{0.2in}
\centerline{\epsffile{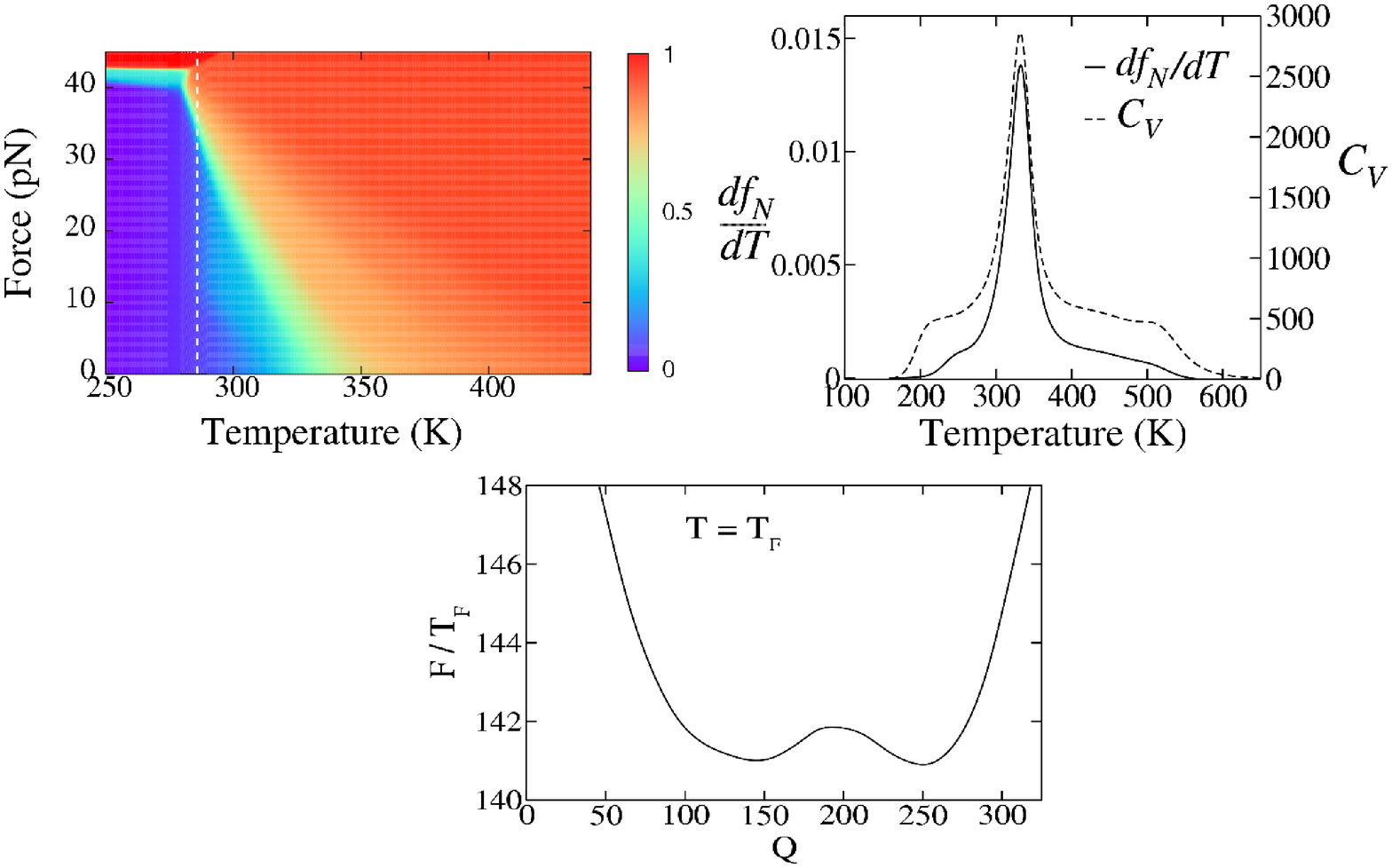}}
\caption{}
\label{diagram}
\end{figure}

%In order to compute thermodynamic quantities we applied the extended
%RE method as described in {\em Materials and Methods}.
The $T-f$ phase diagram
(Fig. \ref{diagram}a) was obtained by monitoring the probability
of being in the native state, $f_N$, as a function of $T$ and $f$, where
$f_N$ is defined as an averaged fraction of native contacts (see Ref.
\onlinecite{MSLi_BJ07} for more details).
The folding-unfolding
transition (the yellow region) is sharp in the low
temperature region, but it becomes less cooperative (the fuzzy
transition region is wider) as $T$ increases.
The folding temperature in the absence of force (peak of $C_v$ or
$df_N/dT$ in Fig. \ref{diagram}b) is equal $T_F=0.64 \epsilon_H/k_B$
which is a bit lower than $T_F=0.67 \epsilon_H/k_B$
of the single Ub \cite{MSLi_BJ07}.
This reflects the fact the folding of the trimer is less cooperative compared
to the monomer due to a small number of native contacts between domains.
One can ascertain this by calculating
the cooperativity index, $\Omega_c$ \cite{Klimov_FD98,Li_PRL04}
for the denaturation transition.
>From simulation data for $df_N/dT$ presented in
Fig. \ref{diagram}b, we obtain
$\Omega_c \approx 40$ which is indeed lower than $\Omega_c \approx 57$ for
the single Ub \cite{MSLi_BJ07} obtained by the same Go model.
According to our previous estimate \cite{MSLi_BJ07},
the experimental value $\Omega_c \approx
384$ is considerably higher than the Go value.
The underestimation of $\Omega _c$  is not only a
shortcoming of the off-lattice Go model \cite{Kouza_JPCA06} but
also a common problem of much more sophisticated force fields in
all-atom models \cite{Phuong_Proteins05}.
Rigid lattice models with side chains give better results for
the cooperativity \cite{MSLi_Physica05,Kouza_JPCA06}
but they are less realistic than the off-lattice ones.
Although the present Go model does not provide the realistic
estimate for cooperativity, it still mimics the experimental
fact, that
folding of a multi-domain protein remains
cooperative observed for not only Ub but also other proteins.

Fig. \ref{diagram}c shows the free energy as a function of native contacts
at $T=T_F$. The folding/unfolding barrier is rather low ($\approx$ 1 kcal/mol),
and is comparable with the case of single ubiquitin \cite{MSLi_BJ07}.
The low barrier is probably an artifact of the simple Go modeling.
The double minimum structure suggests that the trimer is a two-state folder.

\section{Can the force-clamp technique probe refolding pathways of proteins?}

\subsection{Refolding pathways of single ubiquitin}

\begin{figure}
\epsfxsize=6.3in
\vspace{0.2in}
\centerline{\epsffile{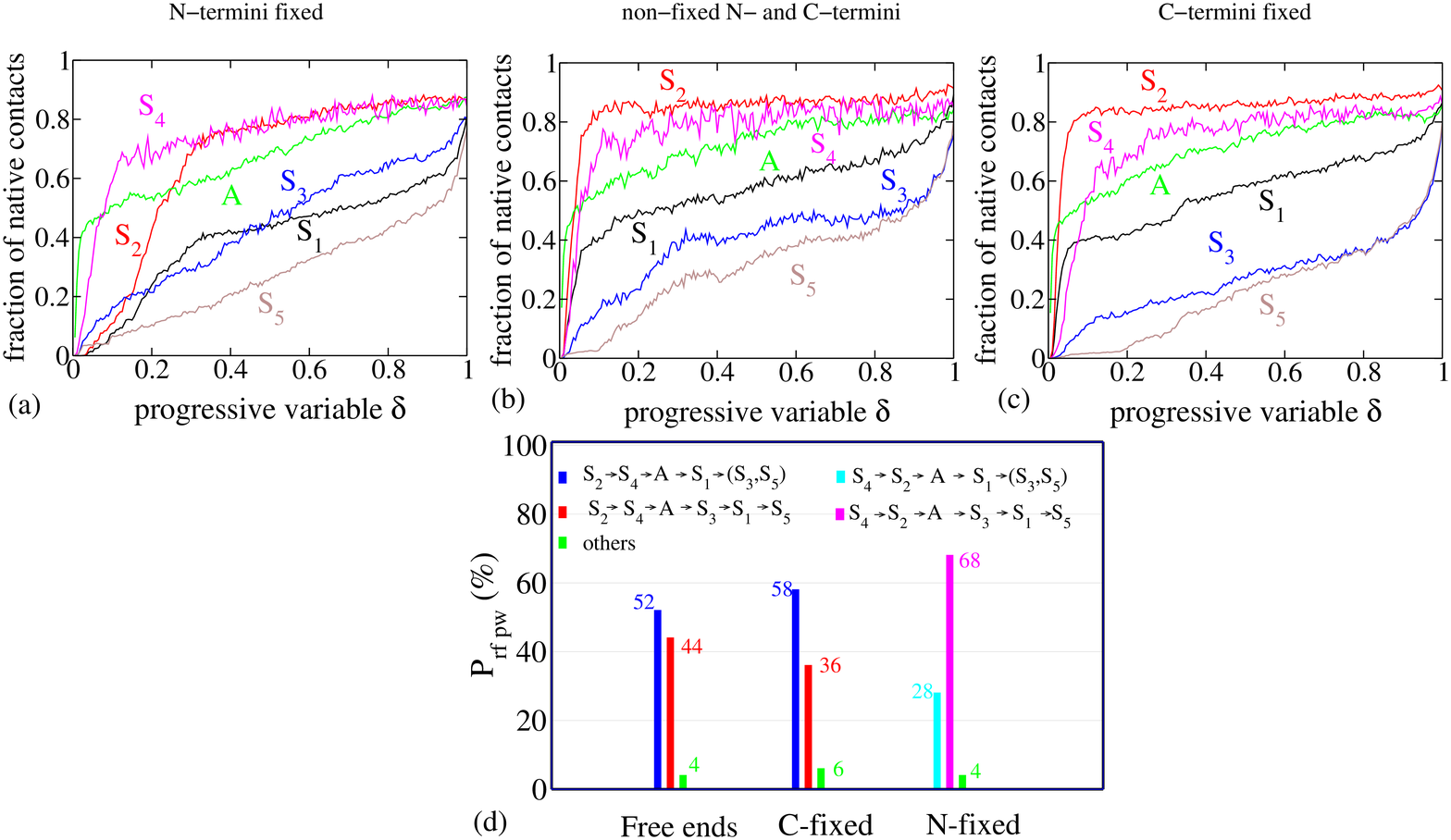}}
\caption{}
\label{single_ub_pathways_fig}
\end{figure}

In order to study refolding under small quenched force we follow the same
protocol as in the experiments \cite{Fernandez_Sci04}.
First, a large force ($\approx 130$ pN) is applied to both termini
to prepare the initial stretched conformations. This force is
then released, but a weak quench force, $f_q$, is applied to study the refolding process.
The refolding  of a single Ub was studied
\cite{MSLi_BJ07,Szymczak_JCP06} in the presence or absence of
the quench force.
Fixing the N-terminal
was found to change the refolding pathways of the free-end
Ub \cite{Szymczak_JCP06}, but the effect of anchoring the C-terminal
has not been studied yet. Here we study this problem in detail, monitoring
the time dependence of native contacts of secondary structures
(Fig. \ref{single_ub_pathways_fig}).
Since the quench force increases the folding time but leaves
the folding pathways unchanged, we present only the results for $f_q=0$
(Fig. \ref{single_ub_pathways_fig}).
Interestingly, the fixed C-terminal and free-end cases have the identical
folding sequencing
\begin{equation}
S2 \rightarrow S4 \rightarrow  A \rightarrow
S1 \rightarrow (S3,S5).
\label{free-end_pathways_eq}
\end{equation}
This is reverse of the unfolding pathway under thermal fluctuations
\cite{MSLi_BJ07}.
As discussed in detail by Li {\em et al.}
\cite{MSLi_BJ07}, Eq. (\ref{free-end_pathways_eq})
 partially agrees with the folding \cite{Went_PEDS05}
and unfolding \cite{Cordier_JMB02} experiments, and simulations
\cite{Fernandez_JCP01,Fernandez_Proteins02,Sorenson_Proteins02}.
Our new finding here is that keeping the C-terminal fixed does not
change the folding pathways.
One should keep in mind that the dominant pathway given by
 Eq. \ref{free-end_pathways_eq}
is valid in the statistical sense.
It occurs in about 52\% and 58\% of events for the free end and C-anchored
cases (Fig. \ref{single_ub_pathways_fig}d), respectively.
The probability of observing an alternate pathway
($S2 \rightarrow S4 \rightarrow  A \rightarrow
S3 \rightarrow S1 \rightarrow S5)$ is $\approx 44$ \% and 36 \% for these
two cases
(Fig. \ref{single_ub_pathways_fig}d). The difference between these two pathways
is only in sequencing  of S1 and S3. Other pathways, denoted in green,
 are also possible
but they are rather minor.

In the case when the N-terminal is fixed (Fig. \ref{single_ub_pathways_fig})
we have the following sequencing
\begin{equation}
S4 \rightarrow S2 \rightarrow A \rightarrow S3 \rightarrow S1 \rightarrow S5
\label{fixedN_pathways_eq}
\end{equation}
which is, in agreement with Ref. \onlinecite{Szymczak_JCP06},
different from the free-end case. We present
folding pathways as the sequencing of
secondary structures,  making comparison with experiments easier
than an approach based on the time
evolution of individual contacts \cite{Szymczak_JCP06}.
The main pathway (Eq. \ref{fixedN_pathways_eq})
occurs in $\approx 68$ \% of events (Fig. \ref{single_ub_pathways_fig}d),
while the competing sequencing $S4 \rightarrow S2 \rightarrow A \rightarrow S1
 \rightarrow (S1, S5)$ (28 \%) and other minor pathways are also possible.
>From Eq. \ref{free-end_pathways_eq} and \ref{fixedN_pathways_eq} it follows
that the force-clamp technique can probe the folding pathways of Ub if one anchores
the C-terminal but not the N-one.

In order to check the robustness of our prediction for refolding pathways
(Eqs. \ref{free-end_pathways_eq} and \ref{fixedN_pathways_eq}),
obtained for the friction $\zeta = 2 \frac{m}{\tau_L}$, we have performed
simulations for the water friction $\zeta = 50 \frac{m}{\tau_L}$
(see II: {\em Model} above). Our
results (not shown) demonstrate that although the folding time
is about twenty times longer compared to the
$\zeta = 2 \frac{m}{\tau_L}$ case, the pathways remain the same.
Thus, within the framework of Go modeling,
the effect of the N-terminus fixation
on refolding pathways of Ub is not an artifact of fast dynamics,
occuring for both large and small friction.
It would be very interesting to verify our prediction
using more sophisticated models. This question is left for future studies.

\subsection{Refolding pathways of three-domain ubiquitin}

\begin{figure}
\epsfxsize=6.3in
\vspace{0.2in}
%\centerline{\epsffile{Q_Rnc_Rg_trimer.eps}}
\centerline{\epsffile{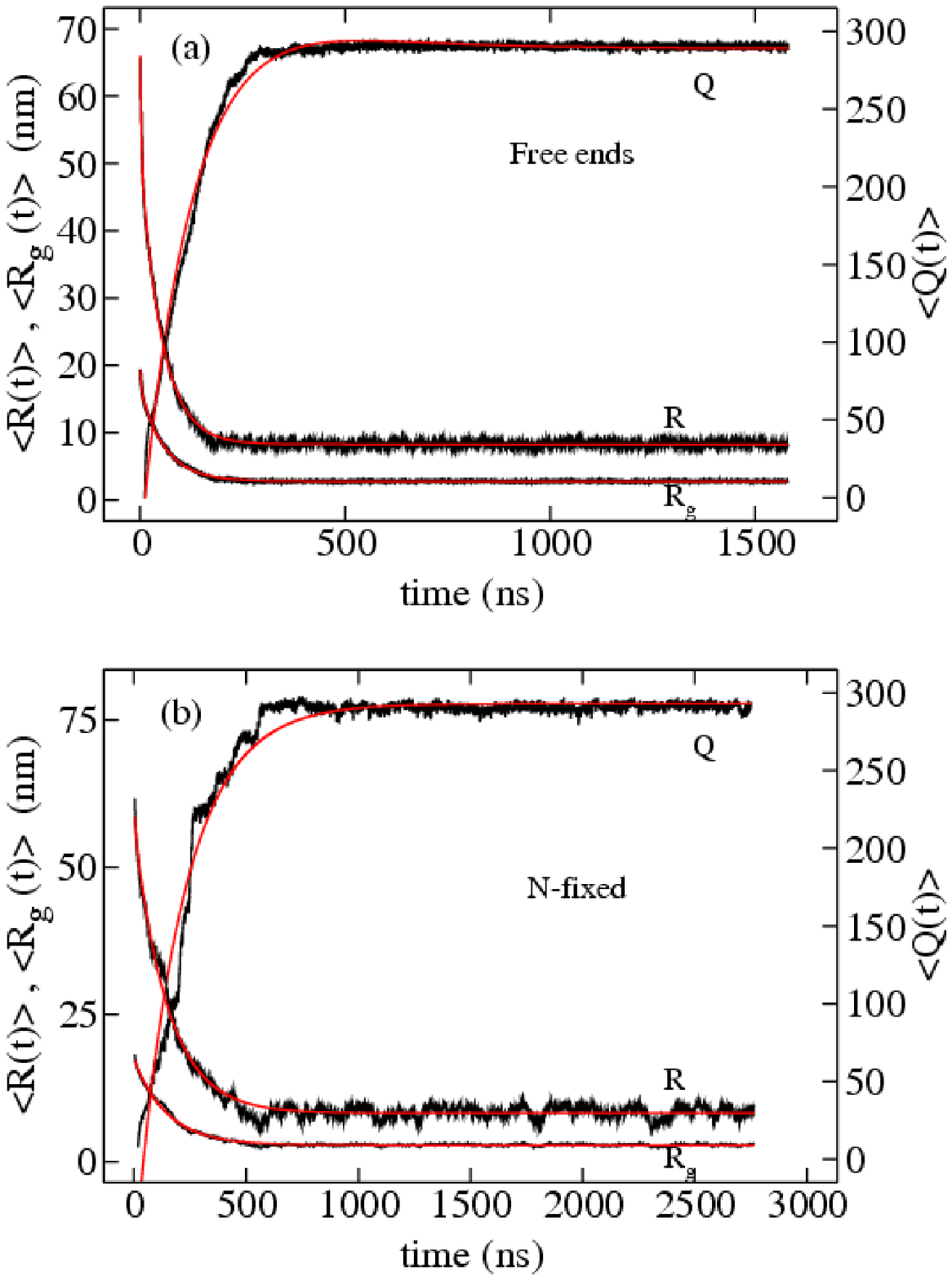}}
\caption{}
\label{Q_Rnc_Rg_trimer_fig}
\end{figure}

The time dependence of the total number of native contacts, $Q$, $R$ and
the  gyration radius, $R_g$, is presented in
Fig. \ref{Q_Rnc_Rg_trimer_fig} for the trimer.
The folding time
$\tau _f \approx$ 553 ns and 936 ns for the free end and N-fixed cases,
respectively. The fact that anchoring one
end slows down refolding by a factor of nearly 2
implies that diffusion-collision processes
\cite{Karplus_Nature76} play an important role in
the Ub folding. Namely, as follows from the diffusion-collision model,
the time required for formation contacts is inversely
proportional to the diffusion coefficient, $D$, of a pair of spherical
units. If one of them is idle, $D$ is halved and
the time needed to form contacts increases accordingly.
The similar effect for unfolding was observed in our recent
work \cite{MSLi_BJ07}.

>From the bi-exponential fitting, we obtain two time scales
for collapsing ($\tau_1$) and compaction ($\tau_2$) where $\tau_1 < \tau_2$.
For $R$, e.g., $\tau_1^R \approx 2.4$ ns and $\tau_2^R \approx 52.3$ ns if
two ends are free, and $\tau_1^R \approx 8.8$ ns and $\tau_2^R \approx 148$ ns
for the fixed-N case. Similar results hold for the time evolution of
$R_g$. Thus, the collapse is much faster than the barrier
limited folding process.
%
% NEW START
Monitoring the time evolution of the end-to-end extension and of
the number of native contacts, one can show (results not shown)
that the refolding of the trimer is staircase-like as observed in the
simulations \cite{Best_Science05,MSLi_BJ07} and the experiments
\cite{Fernandez_Sci04}.
% NEW END

\begin{figure}
\epsfxsize=6.3in
\vspace{0.2in}
%\centerline{\epsffile{fig4.eps}}
\centerline{\epsffile{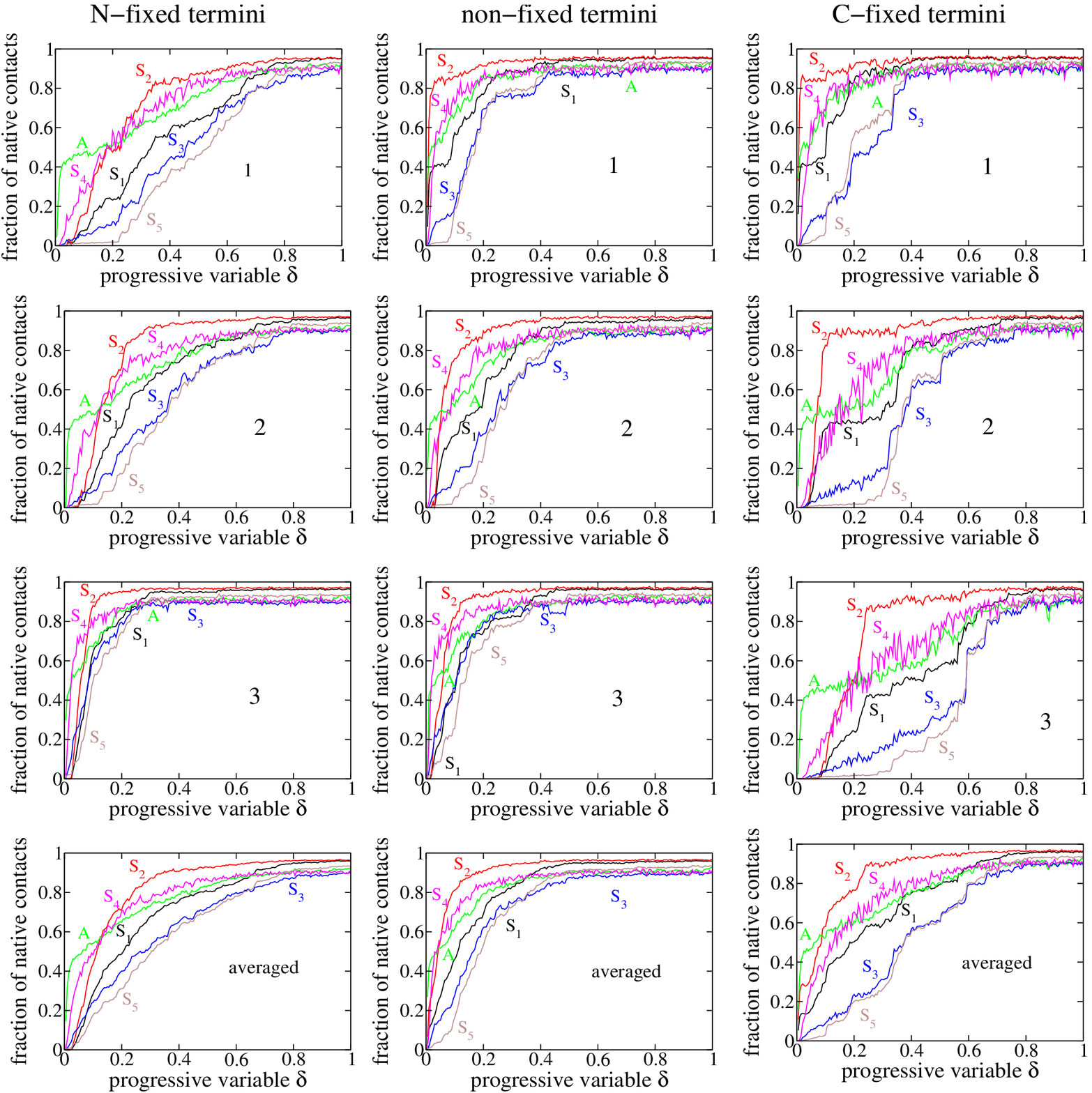}}
\caption{}
\label{trimer_pathways_detail_fig}
\end{figure}

Fig. \ref{trimer_pathways_detail_fig} shows the dependence of the number
of native contacts of the secondary structures of each domain on $\delta$
for three situations: both termini are free and one or the other
of them is fixed.
In each of these cases the folding pathways of three domains
are identical. Interestingly, they are the same,
as given by Eq. \ref{free-end_pathways_eq}, regardless
of we keep one end fixed or not.
As evident from Fig. \ref{trimer_Prfpw_fig},
although the dominant pathway is  the same for three cases its
probabilities are different. It is equal 68\%,
44\% and 43\% for the
C-fixed, free-end and N-fixed cases, respectively. For the last two cases,
the competing pathway
S$_2 \rightarrow$ S$_4 \rightarrow$ A $\rightarrow$ S$_3 \rightarrow$ S$_1 \rightarrow$ S$_5$
 has a reasonably  high
probability of $\approx$ 40\%.

The irrelevance of one-end fixation for refolding
pathways of a multi-domain Ub may be understood
as follows.
Recall that applying the low
quenched force to both termini does not change folding pathways of
single Ub \cite{MSLi_BJ07}. So in the three-domain case,
with the N-end of the first domain fixed,
both termini of the first and second domains are
effectively subjected to external force, and their pathways should remain the
same as in the free-end case. The N-terminal of the third domain is tethered
to the second domain but this would have much weaker effect compared to
the case when it is anchored to a surface. Thus this unit has almost free
ends and its pathways remain unchanged.
Overall, the "boundary" effect
gets weaker as the number of domains becomes
bigger. In order to check this argument, we have performed simulations for
the two-domain Ub. It turns out that the sequencing is roughly the same as
in Fig. \ref{trimer_pathways_detail_fig}, but the common tendency is less
pronounced (results not shown) compared to the trimer case.
Thus we predict that the force-clamp technique can probe
folding pathways of free Ub if one uses either the single domain with the C-terminus
anchored, or the multi-domain construction.

Although fixing one end of the trimer does not influence folding pathways
of individual secondary structures, it affects the folding sequencing
of individual domains (Fig. \ref{trimer_pathways_overall_fig}).
We have the following sequencing $(1,3) \rightarrow 2$,
$3 \rightarrow 2 \rightarrow 1$ and $1 \rightarrow 2 \rightarrow 3$
for the free-end, N-terminal fixed and C-terminal fixed, respectively.
These scenarios are supported by typical snapshots shown in
Fig. \ref{trimer_pathways_overall_fig}. It should be noted that the domain at
the free end folds first in all of three cases in statistical
sense (also true for the two-domain case).
As follows from the bottom of Fig. \ref{trimer_pathways_overall_fig}, if
two ends are free then each of them folds first in about 40 out of 100
observations. The middle unit may fold first, but with much lower probability
of about 15\%. This value remains almost unchanged when one of the ends
is anchored,
and the probability that
the non-fixed unit folds increases to $\ge 80$\%.

%Recently,
%Best and Hummer have studied refolding pathways
%of four-domain Ub \cite{Best_Science05} keeping the N-terminal fixed.
%They found that the folding does not
%commence from the free end but from the third domain. However, this observation
%was demonstrated for one trajectory and it is not clear if it remains valid
%when the results are averaged over many trajectories.

As shown by experiments \cite{Fernandez_Sci04} and simulations
\cite{MSLi_PNAS06,MSLi_BJ07}, one can use the dependence
of refolding times, $\tau _F$,
on the quenched force to find the distance between
the denaturated state and transition state, $x_f$,
along the end-to-end distance
reaction coordinate. Namely, in the Bell approximation
\cite{Bell_Sci78} $\tau_F = \tau_F^0\exp(x_ff_q/k_BT)$, where $\tau_F^0$
is the folding time in the absence of the external force.
Then from Fig. \ref{refold_trimer_fig} we obtain $x_f = 0.74 \pm 0.07$ nm
for the three-domain Ub. Within the error bars this value coincides with
$x_f = 0.96 \pm 0.15$ nm obtained by the same Go model for the single Ub
\cite{MSLi_BJ07}, and with the experimental result $x_f \approx 0.80$ nm
\cite{Fernandez_Sci04}. Our results suggest that the multi-domain structure
leaves  $x_f$ almost unchanged.

\subsection{Is the effect of fixing one terminus on refolding pathways
universal?}

We now ask if the effect of fixing one end
on refolding pathway, observed for Ub, is also valid for other proteins?
To answer this question, we study the single domain I27 from
the muscle protein titin.
We choose this protein as a good candidate
from the conceptual point of view
because its $\beta$-sandwich structure
(see Fig. \ref{titin_str_ref_pathways_fig}a) is very
different from $\alpha/\beta$-structure of Ub.
Moreover, because I27 is subject to
mechanical stress under physiological conditions
\cite{Erickson_Science97}, it is instructive to study
refolding from extended conformations generated by force.
There have been extensive unfolding (see recent review \cite{Sotomayor_Science07}
for
references) and refolding \cite{MSLi_PNAS06} studies
on this system, but the effect of one-end fixation on folding
sequencing of individual secondary structures have not been considered
either theoretically or experimentally.

As follows from Fig. \ref{titin_str_ref_pathways_fig}b,
if two ends are
free then strands A, B and E fold at nearly the same rate.
The pathways of the N-fixed and C-fixed cases are identical,
and they are almost the same as in the free end case
except that the strand A seems to fold after B and E.
Thus, keeping the N-terminus fixed has much weaker effect on the folding
sequencing as compared to the single Ub.
Overall the
effect of anchoring one terminus
has a little effect on the refolding pathways of I27, and
we have the following common sequencing
\begin{equation}
D \rightarrow (B,E) \rightarrow (A,G,A') \rightarrow F \rightarrow C
\end{equation}
for all three cases.
The probability of observing this main pathways varies between 70 and 78\%
(Fig. \ref{titin_str_ref_pathways_fig}e). The second pathway,
D $\rightarrow$ (A,A',B,E,G) $\rightarrow$ (F,C), has considerably lower
probability. Other minor routes to the folded state are also possible.

Because the multi-domain construction weakens this effect, we expect that
the force-clamp spectroscopy can probe refolding pathways for a single and
poly-I27. More importantly, our study reveals that the influence of fixation
on refolding
pathways may depend on the native topology of proteins.

\section{Some problems of unfolding of ubiquitin}

\subsection{Estimation of $x_u$ and the unfolding barrier $\Delta G^{\ddagger}$}

In experiments one usually uses the Bell formula \cite{Bell_Sci78}
\begin{equation}
\tau_{U} = \tau_{U}^0\exp(-x_uf/k_BT)
\label{Bell_eq}
\end{equation}
to extract $x_u$ for two-state proteins from the force dependence
 of unfolding times. Eq. \ref{Bell_eq} is valid if the location of
the transition state does not move under external force.
Recently, assuming that $x_u$ depends on external force
and using the Kramers theory, Dudko {\em et al.} have tried to
go beyond the Bell approximation and proposed
the following force dependence for the unfolding time \cite{Dudko_PRL06}
\begin{equation}
\tau _U \; = \; \tau _U^0
\left(1 - \frac{\nu x_uf}{\Delta G^{\ddagger}}\right)^{1-1/\nu}
\exp\lbrace -\frac{\Delta G^{\ddagger}}{k_BT}[1-(1-\nu x_uf/\Delta G^{\ddagger})^{1/\nu}]\rbrace.
\label{Dudko_eq}
\end{equation}
Here $\Delta G^{\ddagger}$ is the unfolding barrier and $\nu = 1/2$ and 2/3
for the cusp \cite{Hummer_BJ03} and
linear-cubic free energy surface\cite{Dudko_PNAS03}, respectively.
Note that
$\nu =1$ corresponds to the phenomenological
Bell theory (Eq. \ref{Bell_eq}).
One important consequence, following from
Eq. \ref{Dudko_eq}, is that one can apply this technique to estimate not only
$x_u$ but also $G^{\ddagger}$ for $\nu = 1/2$ and 2/3.
This will be done in this section for the single Ub and the trimer.

{\bf A.1. Single Ub}:
Using the Bell approximation, we have already estimated $x_u \approx 2.4$ \AA
\cite{MSLi_BJ07} which is consistent
with the experimental data $x_u = 1.4 - 2.5$ \AA
\cite{Carrion-Vazquez_NSB03,Schlierf_PNAS04,Chyan_BJ04}.
With the help of an all-atom simulation Li {\em et al.} \cite{Li_JCP04}
have shown that $x_u$ does depend on $f$. At low forces,
where the Bell approximation
is valid \cite{MSLi_BJ07}, they
obtained $x_u = 10$ \AA , which is noticeably higher than our and
the experimental value. Presumably,
this is due to the fact
that these authors computed $x_u$ from equilibrium
data, but their sampling was not good enough for such a long protein as Ub.

We now use Eq. \ref{Dudko_eq} with
$\nu = 2/3$ and $\nu = 1/2$ to compute $x_u$ and $\Delta G^{\ddagger}$.
The regions,
where the $\nu = 2/3$ and $\nu = 1/2$ fits works well, are wider  than that for
the Bell scenario (Fig. \ref{refold_unfold_vs_force_fig}). However these fits
can not to cover the entire
force interval. The values of $\tau _U^0, x_u$ and $\Delta G^{\ddagger}$ obtained from
the fitting procedure are listed in Table 1.
In accord with Ref. \onlinecite{Dudko_PRL06}
all of these quantities increase with decreasing $\nu$.
In our opinion, the microscopic theory
($\nu = 2/3$ and $\nu = 1/2$) gives too high a value for
$x_u$ compared to its typical
experimental value \cite{Carrion-Vazquez_NSB03,Schlierf_PNAS04,Chyan_BJ04}.
However, the latter was calculated from fitting experimental
data to the Bell formula,
and it is not clear how much the microscopic theory would change the result.

In order to estimate the unfolding barrier of Ub from the available
experimental data
and compare it with our theoretical estimate, we use the
following formula
\begin{equation}
\Delta G^{\ddagger} = -k_BT\ln(\tau _A/\tau _U^0)
\label{UnfBarrier_eq}
\end{equation}
where $\tau _U^0$ denotes the unfolding time in the absence of force and
$\tau _A$ is a typical unfolding prefactor. Since $\tau _A$ for unfolding is
not known, we use the typical value for folding $\tau _A = 1 \mu$s
\cite{MSLi_Polymer04,Schuler_Nature02}.
Using $\tau _U^0 = 10^4/4$ s
\cite{Khorasanizadeh_Biochem93} and Eq.\ref{UnfBarrier_eq} we obtain
$\Delta G^{\ddagger} =  21.6 k_BT$ which is in reasonable agreement
with our result
$\Delta G^{\ddagger} \approx 17.4 k_BT$, followed from the microscopic fit
with $\nu = 1/2$.
Using the GB/SA contimuum solvation model \cite{Qiu_JPCA97} and the
CHARMM27 force
field \cite{MacKerell_JPCB98}
Li and Makarov \cite{Li_JCP04,Li_JPCB04}
obtained a much
higher
value $\Delta G^{\ddagger} = 29$ kcal/mol $\approx 48.6 k_BT$.
Again,
the large
departure from the experimental result may be related to
poor sampling or to the force filed they used.

{\bf A.2.  The effect of linkage on $x_u$ for single Ub}

One of the most interesting experimental results of
Carrion-Vazquez {\em et al.}\cite{Carrion-Vazquez_NSB03}
is that pulling Ub at different positions changes $x_u$ drastically. Namely,
if the force is applied at the C-terminal and Lys48, then
in the Bell approximation $x_u \approx 6.3$ \AA ,
which is about two and half times larger than the case when the termini N and C
are pulled.
Using the all-atom model
Li and Makarov \cite{Li_JCP04} have shown
that $x_u$ is much larger than 10 \AA.  Thus, a
theoretical reliable estimate for $x_u$ of Lys48-C Ub is not available.
Our aim is to compute $x_u$
employing the present Go-like model \cite{Clementi_JMB00} as
it is successful
in predicting $x_u$ for the N-C Ub.
Fig. \ref{refold_unfold_vs_force_fig} shows the force dependence of
unfolding time of the fragment Lys48-C when the force is
applied to Lys48 and C-terminus. The unfolding time is defined
as the averaged time to stretch this fragment. From the linear fit
($\nu =1$ in Fig. \ref{refold_unfold_vs_force_fig}) at
low forces we obtain $x_u \approx 0.61$ nm which is in good agreement
with the experiment \cite{Carrion-Vazquez_NSB03}.
The Go model is suitable for estimating $x_u$ for not only Ub,
but also for other proteins \cite{MSLi_BJ07a} because
the unfolding is mainly governed by the native topology.
The fact that $x_u$ for the linkage Lys48-C is larger than that of the N-C
Ub may be understood using our recent observation \cite{MSLi_BJ07a}
that it anti-correlates with the contact order (CO)  \cite{Plaxco_JMB98}.
Defining contact formation between any two amino acids ($|i-j| \geq 1$)
as occuring when
the distance between the centers of mass
of side chains $d_{ij} \leq 6.0$ \AA
(see also $http://depts.washington.edu/bakerpg/contact$\_$order/$),
we obtain CO equal 0.075 and 0.15 for the Lys48-C and N-C Ub,
respectively. Thus, $x_u$ of the Lys48-C linkage is larger
than that of the
N-C case because its CO is smaller. This result suggests that
the anti-correlation between $x_u$ and CO may hold
not only when proteins are pulled at termini \cite{MSLi_BJ07a}, but also
when the force is applied to different positions.
Note that the linker (not linkage) effect on $x_u$ has been
studied for protein L \cite{West_PRE06}. It seems
that this effect is
less pronounced compared the effect caused by changing pulling direction
studied here.
We have carried out the microscopic fit for $\nu =1/2$ and $2/3$
(Fig. \ref{refold_unfold_vs_force_fig}). As in the N-C Ub case,
$x_u$ is larger than its Bell value.
However the linkage at Lys48 has a little effect on the activation energy
$\Delta G^{\ddagger}$ (Table 1).

{\bf A.3. Determination of $x_u$ for the three-domain ubiquitin}

Since the trimer is a two-state folder (Fig. \ref{diagram}c),
one can determine
 its averaged distance between the native state and transition state, $x_u$,
along the end-to-end distance reaction coordinate using kinetic
theory \cite{Bell_Sci78,Dudko_PRL06}.
We now ask if the multi-domain structure of Ub changes $x_u$.
As in the
single Ub case \cite{MSLi_BJ07}, there exists a critical force
$f_c \approx 120$pN
separating the low force
and high force regimes (Fig. \ref{refold_unfold_vs_force_fig}).
In the high force region, where the
unfolding barrier disappears, the unfolding time depends on $f$ linearly
(fitting curve not shown) as predicted
theoretically by Evans and Ritchie \cite{Evans_BJ97}.
In the Bell approximation, from the linear fit
(Fig. \ref{refold_unfold_vs_force_fig}) we obtain
$x_u\approx$ 0.24 nm which is exactly
the same as for the single Ub \cite{MSLi_BJ07}.
The multi-domain construction of Ub does not
affect $x_u$, but this may be not the case for other proteins
(M.S. Li, unpublished results).
The values of $\tau _U^0, x_u$ and $\Delta G^{\ddagger}$, extracted
from
the nonlinear fit (Fig. \ref{refold_unfold_vs_force_fig}), are presented
in Table 1. For both $\nu = 1/2$ and $\nu = 2/3$,
$\Delta G^{\ddagger}$ is a bit lower than that
for the single Ub.
In the Bell approximation,
the value of $x_u$ is the same for the single and three-domain Ub but
it is no longer valid for the $\nu = 2/3$ and $\nu = 1/2$ cases.
It would be interesting to perform experiments to check this result and
to see the effect of  multiple domain structure on the free energy landscape.

\subsection{Dependence of unfolding force of single Ub on $T$.}

Recently, using the improved temperature control technique to perform
the pulling experiments
for the single Ub, Yang {\em et al.} \cite{Yang_RSI06}
have found that the unfolding force
depends on $T$ linearly
for 278 K $ \le T \le$ 318 K, and the slope of linear behavior
does not depend on pulling speeds.
 Our goal is
to see if the present Go model
can reproduce this result at least qualitatively, and more importantly,
to check whether the linear dependence holds for the whole temperature
interval where $f_{max} > 0$.

The pulling simulations have been carried at two speeds following the
protocol described
in {\em II: Model}.
Fig. \ref{fmax_T_fig}a shows the force-extension profile of the single
Ub for $T=288$ and 318 K at the pulling speed $v= 4.55\times 10^8$ nm/s.
The peak is lowered as $T$ increases because thermal fluctuations promote
the unfolding of the system. In addition the peak moves toward a
lower extension.
This fact is also understandable, because at higher $T$ a protein can
unfold
at lower extensions due to thermal fluctuations.
For $T=318$ K, e.g., the maximum force is located at the extension
 of $\approx 0.6$ nm, which
corresponds to the plateau observed in the time dependence of
the end-to-end distance under constant force
\cite{Irback_PNAS05,MSLi_BJ07}.
One can show that, in agreement with Chyan {\em et al.}
\cite{Chyan_BJ04}, at this maximum the extension between
strands S$_1$ and S$_5$ is $\approx$ 0.25 nm. Beyond the
maximum, all of the
native contacts between strands S$_1$ and S$_5$ are broken.
At this stage, the chain ends are almost
stretched out, but the rest of the polypeptide chain remains
native-like.

The temperature dependence of the unfolding force, $f_{max}$,
is shown in Fig. \ref{fmax_T_fig}b
for 278 K $\le T \le$ 318 K, and for two pulling speeds.
The experimental results of Yang {\em et al.} are also presented
for comparison. Clearly,
in agreement with experiments \cite{Yang_RSI06}
linear behavior is observed and
the corresponding slopes do not depend on $v$.
Using the fit $f_{max} = f_{max}^0 - \gamma T$ we obtain the ratio
between the simulation and experimental slopes
$\gamma _{sim}/\gamma _{exp} \approx 0.56$.
Thus, the Go model gives
a weaker temperature dependence compared to the experiments.
Given the simplicity of this model, the agreement between theory and experiment
should be considered reasonable, but it would be interesting to check if
a fuller accounting of non-native contacts and environment can improve
our results.

As evident from Fig. \ref{fmax_T_fig}c,
the dependence of $f_{max}$ on $T$ ceases
to be linear for the whole temperature interval.
The nonlinear temperature dependence of $f_{max}$ may be understood
qualitatively using the simple theory of Evans and K. Ritchie
\cite{Evans_BJ97}. Assuming the Bell Eq.\ref{Bell_eq} for the unfolding
time and a linearly ramped force $f = K_rvt$ (see II: {\em Model}) the unfolding
force is given by $f_{max} = \frac{k_BT}{x_u}\ln\frac{K_rvx_u\tau_U^0}{k_BT}$.
A more complicated microscopic
expression for $f_{max}$ is provided in Ref. \onlinecite{Dudko_PRL06}.
Since $\tau_U^0$ is temperature dependent ($x_u$ also displays a weak
temperature dependence \cite{Imparato_PRL07}), the resulting dependence
should be nonlinear.
This result can also be understood by noting that the temperatures considered
here are low enough so that we are not in the entropic limit,
where the linear dependence would be valid for the worm-like model
\cite{Marko_Macromolecules95}.
The arrow in Fig. \ref{fmax_T_fig}c separates two regimes of the $T$-dependence
of $f_{max}$. The crossover takes place roughly in the temperature
interval where the temperature dependence of the equilibrium
critical force changes the slope (Fig. \ref{diagram}).
At low temperatures, thermal fluctuations are weak and
the temperature dependence of $f_{max}$
is weaker compared to the high temperature regime.
Thus the linear dependence observed in the experiments of Yang {\em et al.}
\cite{Yang_RSI06} is valid, but only in the narrow $T$-interval.

\section{Conclusions}

We have adapted the standard temperature RE method to the case when the
force replicas
are considered at a fixed temperature. One can extend the RE method to
cover both temperature and force replicas, as
has been done
for all-atom simulations \cite{Paschek_PRL04} where pressure
is used  instead of force.
One caveat of the force RE method is that the acceptance depends on the
end-to-end distance (Eq. \ref{Delta_eq} and \ref{Metropolis_eq}),
and  becomes inefficient for long proteins.
We can overcome this by increasing the number of replicas,
but this will increase
CPU time substantially. Thus, the question of improving the force RE approach
for long biomolecules remains open.

It has been clearly shown that the secondary structures have the same
folding pathways for all domains, and this is probably the reason why
poly-Ub folds cooperatively.
The folding sequencing of individual domains depends on
whether one end is kept fixed or not.
We predict that the force-clamp technique gives the same folding pathways
for individual secondary structures
as the free-end case, if one anchores either the C-terminal of the single
Ub, or performs the experiments
on the poly-domain Ub. It would be very interesting to verify our
prediction experimentally. Another exciting challenge is to see if
the force-clamp technique
can probe the folding sequencing of other biomolecules using a multi-domain
construction.

We have demonstrated that the anchoring of one terminal has a minor effect on
refolding pathways of I27. This is probably due to the rigidity of
its $\beta$-sandwich structure. It would be interesting to check if this
conclusion holds for other $\beta$-proteins.
The question of to what extent the force-clamp
technique is useful for predicting pathways for $\alpha$-proteins is
left for future studies.

We have shown that, in agreement with the experiment of Carrion-Vazquez
{\em et al.}
\cite{Carrion-Vazquez_NSB03},  the Lys48-C linkage
changes $x_u$ drastically. From the point of view of
biological function,
the linkage Lys63-C is very important, but the study of its mechanical
properties is not as interesting as the Lys48-C because this fragment
is almost stretched out in the native state. Similar to titin and RNA,
$x_u$ and $\Delta G^{\ddagger}$ are sensitive to the parameter
$\nu$, and one has to be very careful in the
interpretation of experimental data.
When comparing theoretical results with experiments, the same
fitting procedure should be used.  The weak inter-domain interaction has
a negligible effect on $x_u$ in the Bell approximation,
but changes become substantial in the nonlinear approximation  ($\nu =1/2$ and
$2/3$). The validity of this observation
for other biomolecules requires further
investigation.
Using the microscopic fitting with $\nu = 1/2$ we have obtained
the reasonable value for $\Delta G^{\ddagger}$ for the single Ub.
This result suggests that the Go modeling is useful for
estimating unfolding barriers of proteins.

Finally, we have reproduced an experiment \cite{Yang_RSI06}
of the linear temperature dependence of unfolding force of Ub
on the quasi-quantitative level.
Moreover, we have shown that for the whole
temperature region the dependence of $f_{max}$ on $T$ is nonlinear,
and the observed linear dependence is valid only for a narrow temperature
interval. This behavior should be common for all proteins because it
reflects the fact that
the entropic limit is not applicable to all temperatures.

MSL thanks D.K. Klimov, G. Morrison, and D. Thirumalai for very
useful discussions. This work was supported by the Polish KBN grant No
1P03B01827, the National Science Council (Taipei) under Grant No.
NSC 96-2911-M 001-003-MY3, Academia Sinica (Taipei) under Grant
No. AS-95-TP-A07, and  National Center for Theoretical Sciences in
Taiwan.

%\bibliographystyle{biophysj}
%\bibliographystyle{proteins}
%%\bibliography{trim}
%\bibliography{Ubiquitin_Li}

\newpage

\newpage
\begin{center}
\begin{tabular}{|c|*{9}{c|}}
\hline
& \multicolumn{3}{c|}{single}& \multicolumn{3}{c|}{Lys48-C}& \multicolumn{3}{c|}{three-domain}\\
\hline
$\nu $& 1/2& 2/3& 1& 1/2 &2/3 &1& 1/2& 2/3 & 1\\
$\quad\tau_U^0 (\mu s)\quad$&13200&1289&9.1&4627&2304&157&1814&756&47\\
$x_{u} (\AA)\quad$ &7.92&5.86&2.4&12.35&10.59&6.1&6.21&5.09&2.4\\
$\quad\Delta G^{\ddagger}(k_BT)\quad$&17.39&14.22&-&15.90&13.94&-&13.49&11.64&-\\
\hline
 \end{tabular}
\end{center}

\vspace{0.5cm}
Table 1. Dependence of $x_u$ on fitting procedures for the three-domain Ub
and Lys48-C. $\nu =1$ corresponds to
the phenomenological Bell approximation
(Eq. \ref{Bell_eq}). $\nu = 1/2$ and 2/3 refer to the
microscopic theory (Eq. \ref{Dudko_eq}).
For comparison we show also
the data for the single Ub which are taken from our previous work
\cite{MSLi_BJ07}.
For the single and three-domain Ub
the force is applied to both termini.

\newpage

\centerline{\Large \bf Figure Captions} \vskip 5 mm

\noindent {\bf FIGURE 1.} (a) Native state conformation of Ub taken from the PDB
(PDB ID: 1ubq). There are five $\beta$-strands: S1 (2-6), S2 (12-16),
S3 (41-45), S4 (48-49) and S5 (65-71), and helix A (23-34).
%(b) Structures B, C, D and E consist of pairs of strands (S1,S2),
%(S1,S5), (S3,S5) and (S3,S4), respectively. In the text
%we also refer to helix A
%as the structure A.
(b) The native conformation of the three-domain
Ub was designed as described in {\em II: Model}.
There are 18 inter-  and 297 intra-domain native contacts.
 \vskip 5 mm

\noindent {\bf FIGURE 2.} (a) The $T-f$ phase diagram obtained by the extended
replica exchange and
histogram method. The force is applied to termini N and C.
The color code for $1-f_N$ is given on the right.
Blue corresponds to the folded state, while red
indicates the unfolded state. The
vertical dashed line denotes to $T=0.85 T_F \approx 285$ K, at which
most of simulations have been performed.
(b) Temperature dependence of the specific heat $C_V$ (right axis) and
$df_N/dT$ (left axis) at $f=0$. Their peaks coincide at $T=T_F$.
%
% NEW START
(c) The dependence of the free energy of the trimer on the total number
of native contacts
$Q$ at $T=T_F$.
% NEW END
\vskip 5mm

\noindent {\bf FIGURE 3.} The dependence of native contacts of $\beta$-strands
and the helix A on the progressive variable $\delta$ when the N-terminal
is fixed (a), both ends are free (b), and C-terminal is fixed
(c). The results are averaged over 200 trajectories.
(d) The probability of refolding pathways in three cases.
each value is written on top of the histograms.

\vskip 5mm
\noindent{\bf FIGURE 4.} (a) The time dependence of $Q$, $R$ and $R_g$ at $T=285$ K for the free end case. (b) The same as in (a) but for the N-fixed case.
The red line is a bi-exponential fit $A(t) = A_0 + a_1\exp(-t/\tau_1)
+ a_2\exp(-t/\tau_2)$.
Results for the C-fixed case are similar to the
$N$-fixed case, and are not shown.

\vskip 5mm

\noindent {\bf FIGURE 5.} The same as in Fig. 3 but for the trimer.
The numbers 1, 2 and 3 refer to the first, second and third domain.
The last row represents the results averaged over three domains.
The fractions of native contacts of each secondary
structure are averaged over 100 trajectories.

\vskip 5mm

\noindent {\bf FIGURE 6.} The probability of different refolding
pathways for the trimer. Each value is shown on top of
the histograms.

\vskip 5mm

\noindent {\bf FIGURE 7.} The dependence of the total number of native
contacts on $\delta$ for the first (green), second (red) and third
(blue) domains. Typical snapshots of the initial, middle and final
conformations for three cases when both two ends are free or one
of them is fixed. The effect of anchoring one terminus
on the folding sequencing of domains is clearly evident.
In the bottom we show the probability of refolding pathways for three
cases. Its value is written on the top of histograms.

\vskip 5mm

\noindent {\bf FIGURE 8.} The dependence of folding times of the trimer
on $f_q$ at $T=285$ K.
$\tau_F$ was computed as the median first passage times of 30-50 trajectories
for each value of $f_q$.
>From the linear fit $y = 6.257 + 0.207x$, we obtained the average distance
 between the unfolded and transition states $x_f = 0.74 \pm 0.07$ nm.

\vskip 5mm

\noindent {\bf FIGURE 9.} (a) Native state conformation of Ig27 domain of titin(PDB ID: 1tit). There are 8 $\beta$-strands: A (4-7), A' (11-15),
B (18-25), C (32-36), D (47-52), E (55-61), F(69-75) and
G (78-88). The dependence of native contacts of different
$\beta$-strands on the progressive variable $\delta$ for the case
when two ends are free (b), the N-terminus is fixed (c) and the
C-terminal is anchored (d).
(e) The probability of observing refolding pathways for three
cases. Each value is written on top of the histograms.

\vskip 5mm

\noindent {\bf FIGURE 10.}
The semi-log plot for the force dependence of unfolding times at $T=285$ K.
Crosses and squares refer the the single Ub and the trimer
with the force applied to N- and C-terminal, respectively.
Circles refer to the
single Ub with the force applied to Lys48 and C-terminal.
Depending on $f$, 30-50 folding events
were using for averaging. In the Bell approximation,
if the N- and C-terminal of the trimer are pulled then
we have the  linear fit $y = 10.448 - 0.066x$ (black line) and the
distance between the native and transition states, $x_u \approx$ 0.24 nm.
The same value of $x_u$ was obtained for the single Ub \cite{MSLi_BJ07}.
In the case when we pull at Lys48 and C-terminal of single Ub the linear fit
(black line) at
low forces is $y = 11.963 - 0.168x$ and $x_u = 0.61$ nm.
The correlation level of fitting is about 0.99.
The red and blue curves correspond to the fits with $\nu =1/2$ and
$2/3$, respectively (Eq. \ref{Dudko_eq}).

\vskip 5mm

\noindent {\bf FIGURE 11.} (a) The force-extension profile obtained at
$T=285$ K (black) and 318 K (red) at the pulling speed
$v= 4.55\times 10^8$ nm/s. $f_{max}$ is located at the extension
$\approx 1$ nm and 0.6 nm for $T=285$ K and 318 K, respectively.
The results
are averaged over 50 independent trajectories.
(b) The dependence of $F_{max}$ on temperature for two values
of $\nu$. The experimental data are taken from
Ref.  \onlinecite{Yang_RSI06} for comparison.
The linear fits for the simulations are
$y = 494.95 - 1.241x$ and $y = 580.69 - 1.335x$. For the experimental sets
we have $y = 811.6 - 2.2x$ and $y = 960.25 - 2.375x$.
(c) The dependence temperature of $F_{max}$ for the whole temperature
region and two values
of $\nu$. The arrow marks the crossover between two nearly linear regimes.
\vskip 5mm

\clearpage

%% FIGURE 1

\clearpage

% FIGURE 2

\clearpage

% FIGURE 3

\clearpage

% FIGURE 4

\clearpage

% FIGURE 5

\clearpage

% FIGURE 6
\begin{figure}
\epsfxsize=6.3in
\vspace{0.2in}
%\centerline{\epsffile{trimer_Prfpw.eps}}
\centerline{\epsffile{fig6.eps}}
\caption{}
\label{trimer_Prfpw_fig}
\end{figure}

\clearpage

% FIGURE 7
\begin{figure}
\epsfxsize=6.3in
%\centerline{\epsffile{fig5.eps}}
\centerline{\epsffile{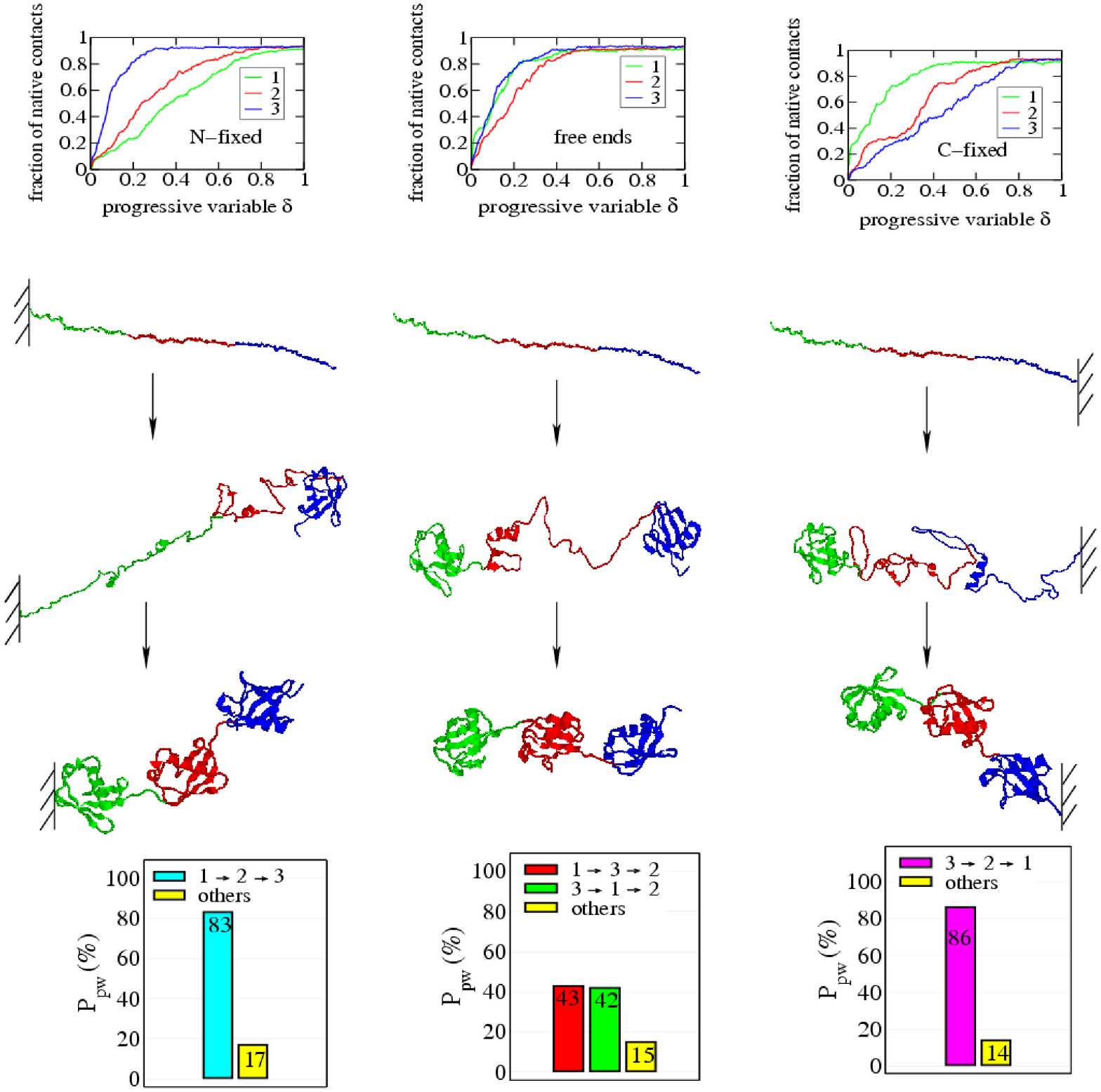}}
\vspace{0.2in}
\caption{}
\label{trimer_pathways_overall_fig}
\end{figure}

\clearpage

% FIGURE 8
\begin{figure}
\epsfxsize=6.3in
\centerline{\epsffile{fig8.eps}}
\vspace{0.2in}
\caption{}
\label{refold_trimer_fig}
\end{figure}

% FIGURE 9
\begin{figure}
\epsfxsize=6.3in
%\centerline{\epsffile{titin_str_ref_pathways.eps}}
\centerline{\epsffile{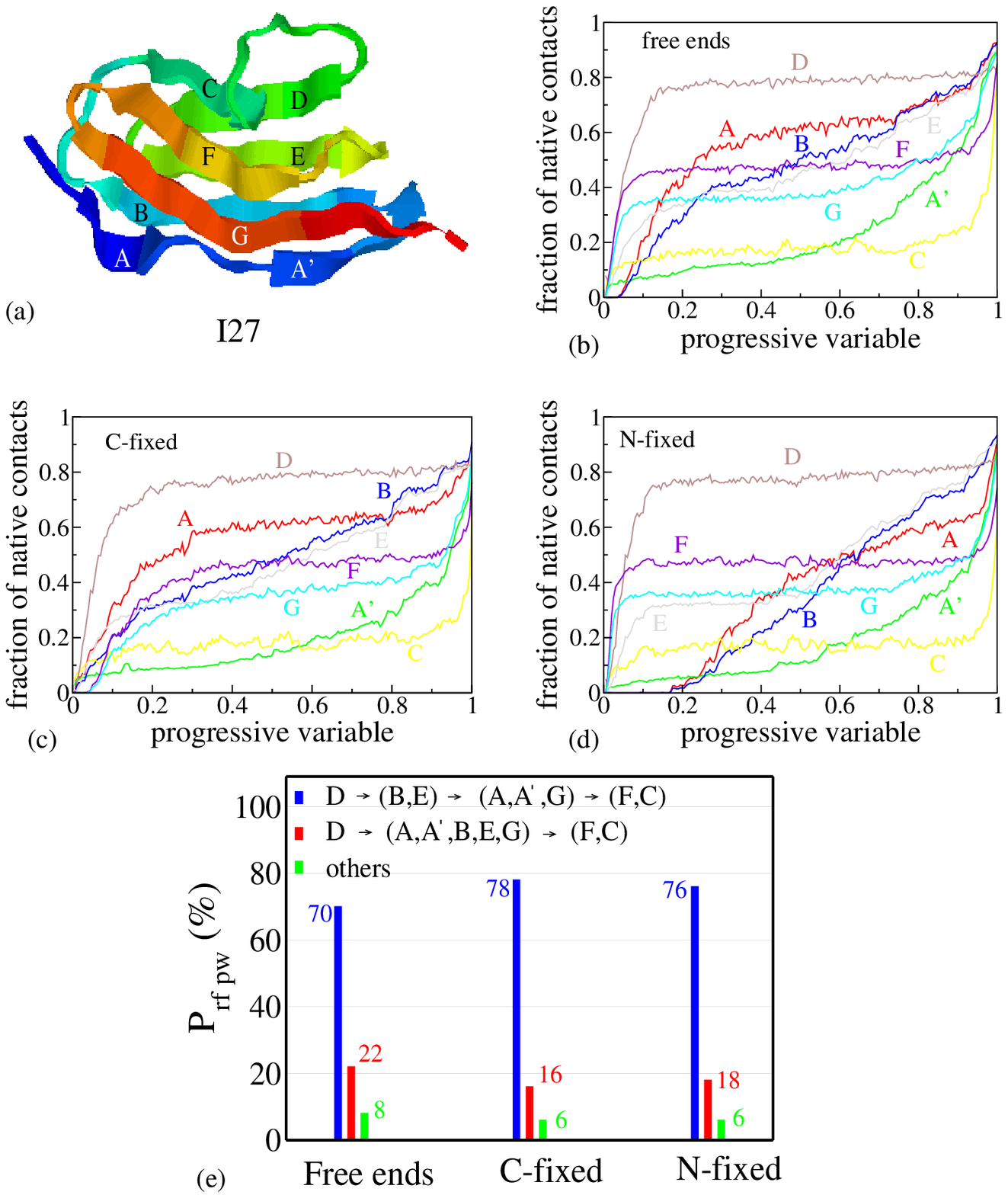}}
\vspace{0.2in}
\caption{}
\label{titin_str_ref_pathways_fig}
\end{figure}

\clearpage

% FIGURE 10
\begin{figure}
\epsfxsize=6.3in
\vspace{0.2in}
%\centerline{\epsffile{fig6.eps}}
\centerline{\epsffile{fig10.eps}}
\caption{}
\label{refold_unfold_vs_force_fig}
\end{figure}

\clearpage

% FIGURE 11
\begin{figure}
\epsfxsize=6.3in
\vspace{0.2in}
%\centerline{\epsffile{fig7.eps}}
\centerline{\epsffile{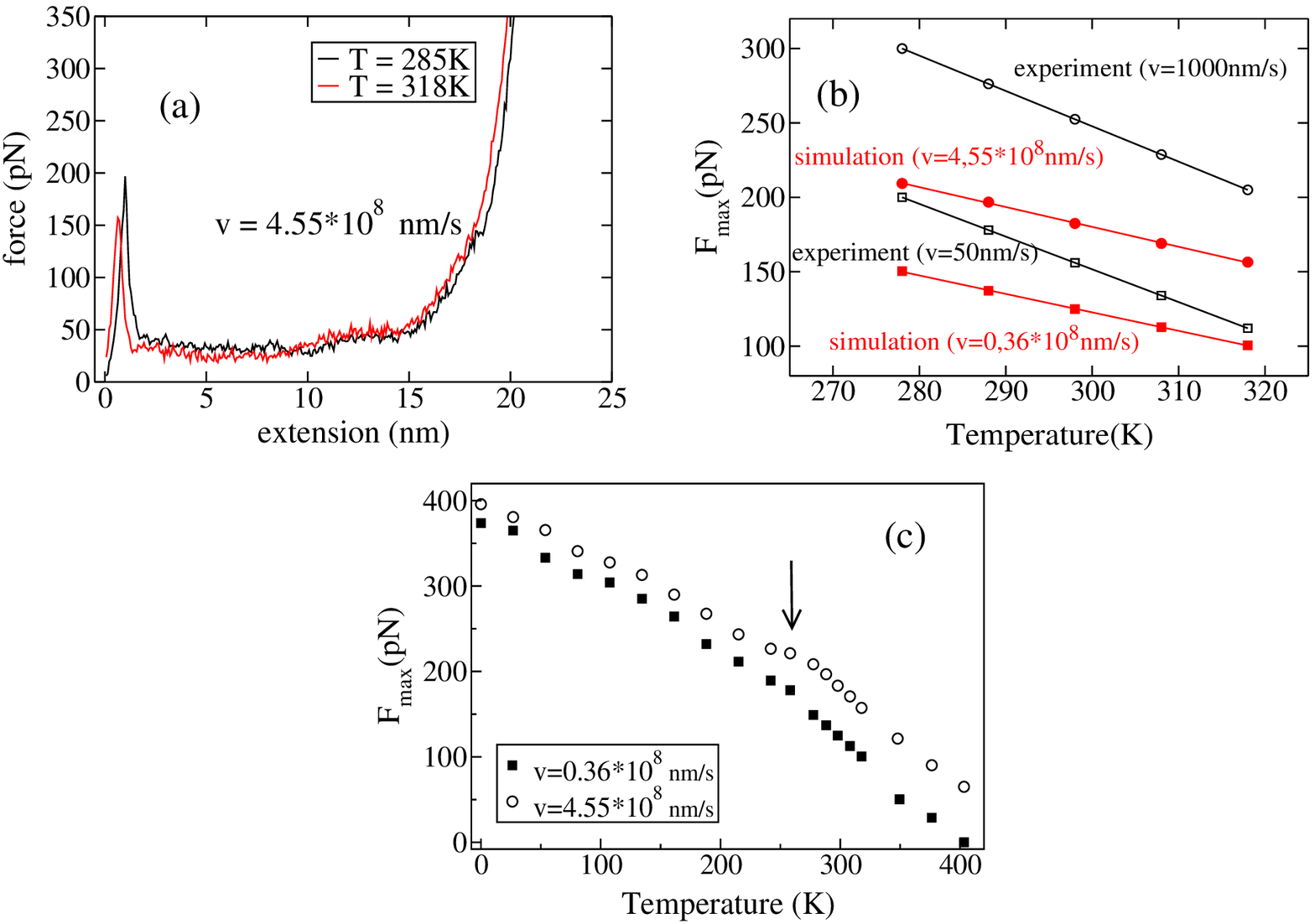}}
\caption{}
\label{fmax_T_fig}
\end{figure}

\clearpage

\end{document}